%% file: ncPRL.tex
\documentclass[prd,twocolumn,showpacs,preprintnumbers,amsmath,amssymb,superscriptaddress]{revtex4}
\usepackage{graphicx}
\usepackage{amsmath}
\usepackage{amssymb}
\usepackage{afterpage}
\usepackage{graphics}
\usepackage{float}
\usepackage{rotating}
\usepackage{xspace}
\usepackage{dcolumn}
\usepackage{bm} 

\newcommand{\nue}{\nu_{e}}
\newcommand{\nus}{\nu_{s}}
\newcommand{\num}{\nu_{\mu}}
\newcommand{\nut}{\nu_{\tau}}

\begin{document}
\preprint{FERMILAB-PUB-08-213-E, arXiv:0807.2424 [hep-ex]}

\title{Search for active neutrino disappearance using neutral-current interactions in the MINOS long-baseline experiment}         

\input{NC-jun08-authors}

\date{\today}          

\begin{abstract}
We report the first detailed comparisons of the rates and spectra of neutral-current neutrino interactions at two widely  separated locations.  A depletion in the rate at the far site would indicate mixing between $\num$ and a sterile particle. No anomalous depletion in the reconstructed energy spectrum is observed.  Assuming oscillations occur at a single mass-squared splitting, a fit to the neutral- and charged-current energy spectra limits the fraction of $\num$ oscillating to a sterile neutrino to be below 0.68 at 90\% confidence level.  A less stringent limit due to a possible contribution to the measured neutral-current event rate at the far site from $\nue$ appearance at the current experimental limit is also presented.
\end{abstract}


\maketitle

Several experiments observing charged-current interactions of neutrinos have provided compelling evidence for $\num$ and $\nue$ disappearance as the neutrinos propagate from the point of production~\cite{previous,superksays,solarprev,kamland,Adamson:2007gu}.  The Super-Kamiokande experiment has reported extensively on the disappearance of $\num$ produced in the atmosphere~\cite{superksays}.  Measurements of solar $\nue$ showed that the disappearance of those neutrinos is due to matter enhanced conversions~\cite{solarprev}.  The KamLAND reactor experiment provided clear evidence for $\overline{\nu}_{e}$ mixing~\cite{kamland}.

These results are conventionally interpreted as mixing among the active neutrino flavors that couple to the electroweak current.  Precise measurements of the $Z$ boson decay width indicate there are only three light active neutrinos~\cite{lep}, but they do not exclude the existence of ``sterile" neutrinos, $\nus$, that do not couple to the electroweak current.  Sterile neutrinos could help resolve several outstanding problems in particle physics and astrophysics.  For example, sterile neutrinos with masses at the eV energy scale can participate in the seesaw mechanism to introduce neutrino masses~\cite{deGouvea:2006gz}  and can also aid in heavy element nucleosynthesis in supernovae~\cite{supernovae}.  The SNO experiment has shown that the total flux of active neutrinos from the Sun agrees with the expectation from solar models~\cite{Ahmad:2002jz}, thereby limiting the extent to which the first or second neutrino mass eigenstates could couple to a sterile neutrino.  While the Super-Kamiokande experiment excludes pure $\num\rightarrow\nus$ and favors pure $\num\rightarrow\nut$ oscillations in its analysis of atmospheric neutrinos, an admixture of the two possibilities is allowed~\cite{superktau} and has attracted considerable attention in the literature~\cite{steriletheory}.

The MINOS experiment has reported a significant deficit of $\num$ at its far detector relative to the near detector through measurement of the rate of $\num$ charged-current (CC) interactions~\cite{Adamson:2007gu,minosCC}.  If this deficit is due solely to conversions of $\num$ to $\nut+\nue$, then the rate of neutral-current (NC) interactions at the far detector remains unchanged from the non-oscillation prediction.  Alternatively, if any  $\num$ convert to a sterile state, then the NC rate would be suppressed and the reconstructed energy spectrum would be distorted.  In this Letter we report the first measurement of the total active neutrino rate using a precisely known long baseline and neutrinos produced with an accelerator.  The reconstructed energy spectra for NC and CC interactions are used to limit the fraction of $\num$ converting to $\nus$ by fitting them to a model of oscillations between $\num, \nut, \nue,$ and $\nus$ dominated by the atmospheric mass-squared splitting.

The neutrino beam is produced using 120~GeV/$c$ protons from the Fermilab Main Injector incident on a graphite target, which is followed by two magnetic focussing horns.  The neutrino energy spectrum can be changed by adjusting the horn current or the position of the target relative to the horns.  The flavor composition of the beam is 92.9\% $\num$, 5.8\% $\overline{\nu}_{\mu}$, and 1.3\% $\nue+\overline{\nu}_{e}$. In this analysis the $\nu$ and $\overline{\nu}$ are assumed to oscillate with the same parameters. The data used in this analysis come from the low energy beam configuration whose peak neutrino energy is 3.3 GeV~\cite{Adamson:2007gu,minosCC}, with an exposure of the far detector to $2.46 \times 10^{20}$ protons on target.

The MINOS near detector is located 1.04~km downstream of the target, has a mass of 0.98~kt, and lies 103~m underground at Fermilab.  The far detector is 734~km downstream of the near detector, has a mass of 5.4~kt, and is located in the Soudan Underground Laboratory in Minnesota, 705~m below the surface.  The fiducial masses used for the near and far detectors are 27~t and 3.8~kt respectively.

The MINOS detectors are steel scintillator tracking calorimeters~\cite{nim}.  The vertically oriented detector planes are composed of 2.54~cm thick steel and 1~cm thick plastic scintillator.  The scintillator layer is comprised of 4.1~cm wide strips with each strip coupled via wavelength-shifting fiber to one pixel of a multi-anode photo-multiplier tube~\cite{Tagg:2004bu,Lang:2005xu}.  The near(far) detector is magnetized to an average toroidal field of 1.3(1.4)~T. 

Hadronic showers resulting from NC interactions generate scintillation light in an average of 12 strips for 1 GeV of deposited energy. Events must have at least 4 strips with signal in order to be considered in the analysis.  Individual scintillator strips are grouped into either reconstructed tracks or showers, which are combined into events.  The vertex for each event is required to be sufficiently far from any edge of the detector to ensure that the final-state hadronic showers are well contained within the fully sampled portion of the detectors.  

The near detector data are used to predict the number of expected events in the far detector, but the ability to make this prediction is complicated by the high rate environment at the near detector.  At an intensity of $2.2\times10^{13}$ protons on target, an average of 16 neutrino interactions are produced in the near detector for each spill~\cite{Adamson:2007gu}.  The reconstruction program separates individual neutrino interactions that occur within the same spill. This initial pass overestimates the number of NC interactions having reconstructed energy, $E_{\text{reco}}$,  $< 1$~GeV by 36\%.  Additional selections making use of event topology and timing are then used to decrease this background.  Events must be separated by at least 40~ns, and events that occur within 120~ns of each other must have verticies separated by at least 1~m in the longitudinal direction~\cite{tobi}.  After applying these criteria, the remaining background from poorly reconstructed events with $E_{\text{reco}} < 1$~GeV is 7\%.

The rate of neutrino interactions from the neutrino beam in the far detector is much lower than in the near detector, with approximately 1 interaction for every $10^{4}$ spills.  Interactions from the beam neutrinos are identified using a window around the GPS time stamp of the spills of $-2~\mu\text{s} < t < 12~\mu$s where $t=0$ is the expected start time at the far detector of the $10~\mu$s spill.  Given the low rate of neutrino interactions in the far detector, spurious events that are coincident with the beam spills from noise, cosmic-ray muons, or poor event reconstruction can introduce backgrounds to the analysis.  Additional criteria are used to remove such events, leaving a residual background of $< 1\%$ of the signal~\cite{phill}.  

\begin{figure}
\centerline{\includegraphics[width=3.3in]{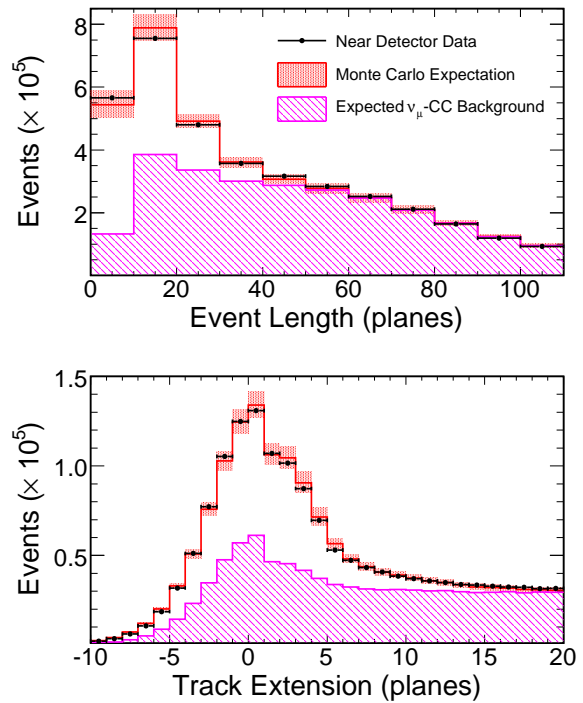}}
\caption{Distributions of event-topology parameters used to separate NC-like from CC-like events. Data from the near detector (solid points) are shown superposed on the total Monte Carlo expectation.  The hatched distribution shows the $\num$-CC background as determined by the Monte Carlo simulation.  The systematic uncertainty for the Monte Carlo expectation is shown by the shaded band.}
\label{fig:cut_parameters}
\end{figure}

\begin{figure}
\centerline{\includegraphics[width=3.3in]{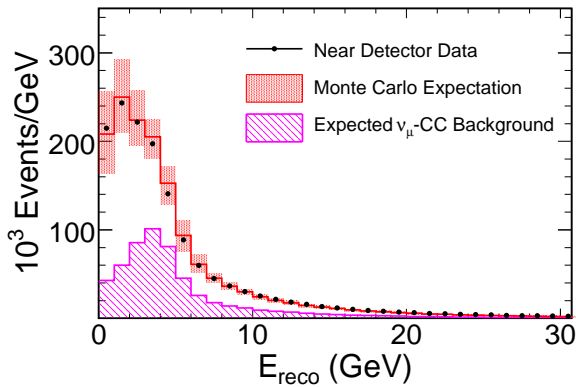}}
\caption{The reconstructed energy spectrum for NC-like events in the near detector.  The data (solid points) and the Monte Carlo expectation including systematic uncertainties (solid histogram with shaded band) are shown.  The hatched distribution shows the expected $\num$-CC background.}
\label{fig:near_spect}
\end{figure}

Charged-current interactions are identified by the presence of a track that may or may not be associated with a shower.  Neutral-current interactions typically have a single hadronic shower, although the reconstruction may identify a track in the event; such tracks could come from pions, but are mostly reconstruction artifacts.  An event is classified as NC-like if it has a reconstructed shower, is shorter than 60 planes, and has no track extending more than 5 consecutive planes beyond the shower~\cite{osiecki}.  Distributions of these event-topology parameters for near detector events are shown in Fig.~\ref{fig:cut_parameters}.  
The principal background in the spectrum of NC-like events comes from highly inelastic $\num$-CC interactions.  The $E_{\text{reco}}$ spectrum of NC-like events in the near detector is shown in Fig.~\ref{fig:near_spect}.  The distributions in Figs.~\ref{fig:cut_parameters}~and~\ref{fig:near_spect} show good agreement between the data and Monte Carlo simulation.   

The Monte Carlo simulation is used to make an initial estimate of the ratio of event yields in the far and near detectors as a function of $E_{\text{reco}}$.  This ratio is multiplied by the observed energy spectrum in the near detector to produce a far detector prediction of the NC-like event spectrum.  The true energy of the simulated neutrinos in each reconstructed energy bin of the prediction is used to determine the effect of oscillations for that range of reconstructed energy.  To avoid biases, the methods for identifying NC-like events and predicting the far detector spectrum were developed and tested using only the near detector data and Monte Carlo simulation, and the analysis procedures were finalized prior to examining data in the far detector.  

\begin{figure}
\centerline{\includegraphics[width=3.3in]{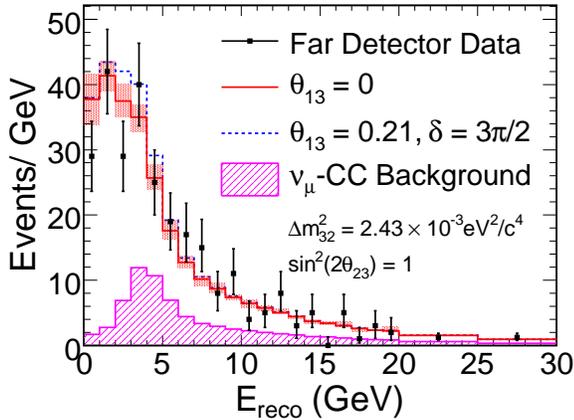}}
\caption{Spectrum of observed NC-like events in the far detector with predictions for the two oscillation hypotheses described in the text. The filled regions in each bin indicates the systematic uncertainty in the predicted rates.}
\label{fig:3flavor_comps}
\end{figure}

Figure~\ref{fig:3flavor_comps} shows the measured and predicted $E_{\text{reco}}$ spectra at the far detector.  The spectra are compared using a statistic, $R$, which expresses the agreement between the predicted and observed number of events in the far detector:  
\begin{equation}
R \equiv \frac{N_{\text{Data}}-B_{\text{CC}}}{S_{\text{NC}}},
\label{eq:fdef}
\end{equation}
where, within a given energy range, $N_{\text{Data}}$ is the measured event count, $B_{\text{CC}}$ is the extrapolated CC background from all flavors, and $S_{\text{NC}}$ is the extrapolated number of NC interactions. The values of $S_{\text{NC}}$ and contributions to $B_{\text{CC}}$ are calculated in the framework of three neutrino oscillations and are shown in Table~\ref{table:nums}.  Because the disappearance of $\num$ occurs mainly for true neutrino energies $< 6$~GeV~\cite{minosCC}, the data are separated into two samples.  Events with $E_{\text{reco}}<3$~GeV are grouped into a low-energy sample while events with $3~\text{GeV}<E_{\text{reco}}<120$~GeV are grouped into a high-energy sample.  The median true neutrino energies of the low and high energy samples are  3.1~GeV and 7.9~GeV respectively.  The values of $R$ calculated for these ranges in $E_{\text{reco}}$ are shown in Table~\ref{table:nums}.  In the region with $E_{\text{reco}} < 3$~GeV, $R$ differs from 1 by $1.3\sigma$.  Over the full energy range, $0-120$~GeV, the depletion of the total NC event rate is limited to be below 17\% at 90\% confidence level.
\begin{table}
\caption{\label{table:nums} Values of $N_{\text{Data}}, S_{\text{NC}},$ and the contributions to $B_{\text{CC}}$ for various reconstructed energy ranges.  Also shown are the values of $R$.  The numbers in parentheses are calculated including $\nue$ appearance at the upper limit discussed in the text.}  
\begin{tabular}{llcccc}
\hline
$E_{\text{reco}}$ (GeV)      & $N_{\text{Data}}$ & ~~$S_{\text{NC}}$~~ & ~~$B^{\num}_{\text{CC}}$~~ & ~~$B^{\nut}_{\text{CC}}$~~ & ~~$B^{\nue}_{\text{CC}}$~~ \\ \hline
$0-3$       &100&101.1&11.2&1.0&1.8~~(9.3)   \\
$3-120$  &191&98.0&64.2&3.5&11.8 (24.6) \\
\hline 
$0-3$ &\multicolumn{5}{l}{$R=0.85\pm0.10\pm0.07~~(0.78\pm0.10\pm0.07)$} \\
$3-120$ &\multicolumn{5}{l}{$R=1.14\pm0.14\pm0.10~~(1.02\pm0.14\pm0.10)$} \\
$0-120$ &\multicolumn{5}{l}{$R=0.99\pm0.09\pm0.07~~(0.90\pm0.09\pm0.08)$} \\
\hline 
\end{tabular}
\end{table}

The principal sources of systematic uncertainty in $R$ are listed in Table~\ref{table:syst}.  The absolute scale of the hadronic energy is known to within 12\%,  of which 10\% reflects uncertainties in the final-state interactions in the nucleus and 6\% results from uncertainty in the detector response to single hadrons.  The relative calibration of the hadronic energy between the two detectors has an uncertainty of 3\%~\cite{Adamson:2007gu}, and the relative normalization between the detectors has an uncertainty of 4\%. The uncertainty in the near detector event count due to the selection criteria is 15\% for $E_{\text{reco}} < 0.5$~GeV;  3\% for events with  $0.5~\text{GeV} < E_{\text{reco}} < 1$~GeV; and is negligible for $E_{\text{reco}} > 1$~GeV. The effect of these uncertainties on $R$ is shown in Table~\ref{table:syst}. 

The uncertainty on the size of the $\num$-CC background was determined by comparing the near detector NC-like reconstructed energy spectrum from the low energy beam configuration used in this analysis with the spectra from three other beam configurations with higher average neutrino energy.  In each reconstructed energy bin, $i$, of the low energy beam the total number of events is the sum of the NC and CC interactions, $N_{i} = \text{NC}_{i} + \text{CC}_{i}$. The quantity $r_{i}^{\text{NC}}(r_{i}^{\text{CC}})$ is defined as the ratio of the number of NC(CC) interactions in each energy bin in an alternative beam configuration to the corresponding number in the low energy beam configuration. The value of $\text{CC}_{i}$ can be calculated from the spectrum in another beam,
\begin{equation}
\text{CC}_{i} = \frac{r^{\text{NC}}_{i}N_{i} - N^{A}_{i}}{r^{\text{NC}}_{i}-r^{\text{CC}}_{i}},
\end{equation}
where $N_{i}^{A}$ is the total number of events observed in the alternate beam configuration. The values of $r_{i}^{\text{NC}}$ and $r_{i}^{\text{CC}}$ are taken from the Monte Carlo simulation. The uncertainty in the $\num$-CC background is taken as the difference between the uncertainty-weighted average value of $\text{CC}_{i}$ measured using the different beam configurations and the value predicted by the Monte Carlo simulation. That difference is consistent within 15\% for all reconstructed energies.  The size of the $\num$-CC background at the far detector depends on the parameters for $\nu_{\mu}\rightarrow\nu_{\tau}$ oscillations used in the prediction.  The MINOS measured values of $\Delta m^{2}_{32} = 2.43\times10^{-3}$~eV$^{2}/c^{4}$ and $\theta_{23} = \pi/4$~\cite{minosCC} were used for the prediction, and variations within the $1\sigma$ range of these parameters change the $\num$-CC background in the far detector by less than $10$\%.  
\begin{table}
\caption{\label{table:syst} Sources of systematic uncertainties considered in this analysis and their effect on $R$.}
\begin{tabular}{lll}
\hline
& 0 -- 3~GeV & ~~~~~3 -- 120~GeV \\
\hline
Absolute $E_{\text{had}}$ & $\pm<0.01$ & ~~~~~$\pm0.05$\\ 
Relative $E_{\text{had}}$ & $\pm0.03$ & ~~~~~$\pm0.04$ \\ 
Normalization & $\pm0.04$ & ~~~~~$\pm0.08$ \\ 
Near detector selection & $\pm0.02$ & ~~~~~-- \\ 
$\nu_\mu$-CC background & $\pm0.03$ & ~~~~~$\pm 0.01$ \\ 
\hline
Total: & $\pm0.07$ & ~~~~~$\pm0.10$ \\ 
\hline 
\end{tabular}
\end{table}

Because the selection criteria identify $\nue$-CC interactions as NC-like with nearly 100\% efficiency, the background from $\nue$ inherent in the beam and $\nu_{\mu}\rightarrow\nu_{e}$ oscillations is also considered.  An upper limit for the $\nue$-CC rate in the far detector was estimated using the normal mass hierarchy with $\theta_{12} = 0.61$~rad,  $\theta_{13} = 0.21$~rad, $\delta = 3\pi/2$~rad, $\Delta m^{2}_{21} = 7.59\times 10^{-5}$eV$^{2}/c^{4}$, and $\Delta m^{2}_{32} = 2.43\times10^{-3}$~eV$^{2}/c^{4}$~\cite{kamland,minosCC}.  The choice of $\theta_{13}$ corresponds to the 90\% confidence level upper limit for the chosen $\Delta m^{2}_{32}$ value~\cite{Apollonio:2002gd}.   The contribution to $B_{\text{CC}}$ from $\nue$ and the values of $R$ in the different energy ranges under these assumptions are shown in Table~\ref{table:nums}.  

The data shown in Fig.~\ref{fig:3flavor_comps} can be combined with the data from CC interactions to determine whether the previously observed $\num$ disappearance is due solely to oscillations between the active neutrinos, or if oscillations between active and sterile neutrinos also occur.  To determine the fraction of $\num$ that have converted to a sterile state, the data are fit to a model that assumes oscillations between $\num,\nut,$ and $\nus$ occur at a single mass-squared splitting.  The probabilities for $\num$ to remain $\num$ or convert to $\nus$ are
\begin{eqnarray}
\nonumber
P_{\num\rightarrow\num} & = &1 - \alpha_{\mu}\sin^{2}(1.27\Delta m^{2}L/E), \text{and} \label{eq:probsparam}
\\ 
P_{\num\rightarrow\nus}  & = & \alpha_{s}\sin^{2}(1.27\Delta m^{2}L/E), \\ \nonumber
\end{eqnarray}
where $\Delta m^{2}$ is the atmospheric mass-squared splitting in eV$^{2}/c^{4}$, $L=735$~km, $E$ is the energy of the neutrino in GeV, and $\alpha_{\mu}$ and $\alpha_{s}$ are phenomenological parameters related to the mixing angles.  A simultaneous fit to the $E_{\text{reco}}$ spectrum in Fig.~\ref{fig:3flavor_comps} and the $\num-$CC energy spectrum yields the energy independent fraction of $\num$ that oscillate to $\nus$, 
\begin{equation}
 f_{s}\equiv\frac{P_{\num\rightarrow\nus}}{1-P_{\num\rightarrow\num}} = 0.28^{+0.25}_{-0.28}(\text{stat.+syst.}),
\end{equation}
with $\chi^{2}=46.5$ for 43 degrees of freedom and $f_{s}< 0.68$ at 90\% confidence level.  The fit includes the systematic uncertainties in Table~\ref{table:syst} as nuisance parameters.  Including electron neutrino appearance at the previously discussed upper limit results in $f_{s} = 0.43^{+0.23}_{-0.27}(\text{stat.+syst.})$ with $\chi^{2} = 46.6$ and $f_{s} < 0.80$ at 90\% confidence level.

In summary, we have reported the first measurements of neutrino neutral-current rates and spectra in an accelerator long baseline neutrino experiment. The rates at the near and far detectors are consistent with expectations from decay kinematics and geometry, providing new support for the interpretation of muon neutrino disappearance as oscillations among the three active neutrinos. This result provides the best limits to date on the fraction of muon neutrinos which may convert to sterile neutrinos in oscillations associated with the atmospheric mass-squared splitting.

This work was supported by the US DOE, the UK STFC, the US NSF, the State and University of Minnesota, the University of Athens, Greece, and Brazil's FAPESP and CNPq.  We thank S. Parke for useful discussions.  We are grateful to the Minnesota Department of Natural Resources, the crew of the Soudan Underground Laboratory, and the staff of Fermilab for their contribution to this effort.

\bibliography{ncPRL}

\end{document}

%% file: NC-jun08-authors.tex
\newcommand{\Cambridge}{Cavendish Laboratory, University of Cambridge, Madingley Road, Cambridge CB3 0HE, United Kingdom}
\newcommand{\FNAL}{Fermi National Accelerator Laboratory, Batavia, Illinois 60510, USA}
\newcommand{\RAL}{Rutherford Appleton Laboratory, Chilton, Didcot, Oxfordshire, OX11 0QX, United Kingdom}
\newcommand{\UCL}{Department of Physics and Astronomy, University College London, Gower Street, London WC1E 6BT, United Kingdom}
\newcommand{\Caltech}{Lauritsen Laboratory, California Institute of Technology, Pasadena, California 91125, USA}
\newcommand{\ANL}{Argonne National Laboratory, Argonne, Illinois 60439, USA}
\newcommand{\Athens}{Department of Physics, University of Athens, GR-15771 Athens, Greece}
\newcommand{\NTUAthens}{Department of Physics, National Tech. University of Athens, GR-15780 Athens, Greece}
\newcommand{\Benedictine}{Physics Department, Benedictine University, Lisle, Illinois 60532, USA}
\newcommand{\BNL}{Brookhaven National Laboratory, Upton, New York 11973, USA}
\newcommand{\CdF}{APC -- Universit\'{e} Paris 7 Denis Diderot, 10, rue Alice Domon et L\'{e}onie Duquet, F-75205 Paris Cedex 13, France}
\newcommand{\Cleveland}{Cleveland Clinic, Cleveland, Ohio 44195, USA}
\newcommand{\Delhi}{Department of Physics and Astrophysics, University of Delhi, Delhi 110007, India}
\newcommand{\GEHealth}{GE Healthcare, Florence South Carolina 29501, USA}
\newcommand{\Harvard}{Department of Physics, Harvard University, Cambridge, Massachusetts 02138, USA}
\newcommand{\HolyCross}{Holy Cross College, Notre Dame, Indiana 46556, USA}
\newcommand{\IIT}{Physics Division, Illinois Institute of Technology, Chicago, Illinois 60616, USA}
\newcommand{\Indiana}{Indiana University, Bloomington, Indiana 47405, USA}
\newcommand{\ITEP}{High Energy Experimental Physics Department, ITEP, B. Cheremushkinskaya, 25, 117218 Moscow, Russia}
\newcommand{\JMU}{Physics Department, James Madison University, Harrisonburg, Virginia 22807, USA}
\newcommand{\LASL}{Nuclear Nonproliferation Division, Threat Reduction Directorate, Los Alamos National Laboratory, Los Alamos, New Mexico 87545, USA}
\newcommand{\Lebedev}{Nuclear Physics Department, Lebedev Physical Institute, Leninsky Prospect 53, 119991 Moscow, Russia}
\newcommand{\LLL}{Lawrence Livermore National Laboratory, Livermore, California 94550, USA}
\newcommand{\MIT}{Lincoln Laboratory, Massachusetts Institute of Technology, Lexington, Massachusetts 02420, USA}
\newcommand{\Minnesota}{University of Minnesota, Minneapolis, Minnesota 55455, USA}
\newcommand{\Crookston}{Math, Science and Technology Department, University of Minnesota -- Crookston, Crookston, Minnesota 56716, USA}
\newcommand{\Duluth}{Department of Physics, University of Minnesota -- Duluth, Duluth, Minnesota 55812, USA}
\newcommand{\Oxford}{Subdepartment of Particle Physics, University of Oxford, Oxford OX1 3RH, United Kingdom}
\newcommand{\Pittsburgh}{Department of Physics and Astronomy, University of Pittsburgh, Pittsburgh, Pennsylvania 15260, USA}
\newcommand{\IHEP}{Institute for High Energy Physics, Protvino, Moscow Region RU-140284, Russia}
\newcommand{\RoyalH}{Physics Department, Royal Holloway, University of London, Egham, Surrey, TW20 0EX, United Kingdom}
\newcommand{\Carolina}{Department of Physics and Astronomy, University of South Carolina, Columbia, South Carolina 29208, USA}
\newcommand{\SLAC}{Stanford Linear Accelerator Center, Stanford, California 94309, USA}
\newcommand{\Stanford}{Department of Physics, Stanford University, Stanford, California 94305, USA}
\newcommand{\StJohnFisher}{Physics Department, St. John Fisher College, Rochester, New York 14618 USA}
\newcommand{\Sussex}{Department of Physics and Astronomy, University of Sussex, Falmer, Brighton BN1 9QH, United Kingdom}
\newcommand{\TexasAM}{Physics Department, Texas A\&M University, College Station, Texas 77843, USA}
\newcommand{\Texas}{Department of Physics, University of Texas at Austin, 1 University Station C1600, Austin, Texas 78712, USA}
\newcommand{\TechX}{Tech-X Corporation, Boulder, Colorado 80303, USA}
\newcommand{\Tufts}{Physics Department, Tufts University, Medford, Massachusetts 02155, USA}
\newcommand{\UNICAMP}{Universidade Estadual de Campinas, IF-UNICAMP, CP 6165, 13083-970, Campinas, SP, Brazil}
\newcommand{\USP}{Instituto de F\'{i}sica, Universidade de S\~{a}o Paulo,  CP 66318, 05315-970, S\~{a}o Paulo, SP, Brazil}
\newcommand{\Warsaw}{Department of Physics, Warsaw University, Ho\.{z}a 69, PL-00-681 Warsaw, Poland}
\newcommand{\Washington}{Physics Department, Western Washington University, Bellingham, Washington 98225, USA}
\newcommand{\WandM}{Department of Physics, College of William \& Mary, Williamsburg, Virginia 23187, USA}
\newcommand{\Wisconsin}{Physics Department, University of Wisconsin, Madison, Wisconsin 53706, USA}
\newcommand{\deceased}{Deceased.}

\affiliation{\ANL}
\affiliation{\Athens}
\affiliation{\Benedictine}
\affiliation{\BNL}
\affiliation{\Caltech}
\affiliation{\Cambridge}
\affiliation{\UNICAMP}
\affiliation{\CdF}
\affiliation{\FNAL}
\affiliation{\Harvard}
\affiliation{\IIT}
\affiliation{\Indiana}
\affiliation{\ITEP}
\affiliation{\Lebedev}
\affiliation{\LLL}
\affiliation{\UCL}
\affiliation{\Minnesota}
\affiliation{\Duluth}
\affiliation{\Oxford}
\affiliation{\Pittsburgh}
\affiliation{\RAL}
\affiliation{\USP}
\affiliation{\Carolina}
\affiliation{\Stanford}
\affiliation{\Sussex}
\affiliation{\TexasAM}
\affiliation{\Texas}
\affiliation{\Tufts}
\affiliation{\Warsaw}
\affiliation{\Washington}
\affiliation{\WandM}

\author{P.~Adamson}
\affiliation{\FNAL}

\author{C.~Andreopoulos}
\affiliation{\RAL}

\author{K.~E.~Arms}
\affiliation{\Minnesota}

\author{R.~Armstrong}
\affiliation{\Indiana}

\author{D.~J.~Auty}
\affiliation{\Sussex}


\author{D.~S.~Ayres}
\affiliation{\ANL}

\author{C.~Backhouse}
\affiliation{\Oxford}

\author{B.~Baller}
\affiliation{\FNAL}



\author{G.~Barr}
\affiliation{\Oxford}

\author{W.~L.~Barrett}
\affiliation{\Washington}


% ...

\collaboration{The MINOS Collaboration}
\noaffiliation

%% file: ncPRL.bbl
\begin{thebibliography}{19}
\expandafter\ifx\csname natexlab\endcsname\relax\def\natexlab#1{#1}\fi
\expandafter\ifx\csname bibnamefont\endcsname\relax
  \def\bibnamefont#1{#1}\fi
\expandafter\ifx\csname bibfnamefont\endcsname\relax
  \def\bibfnamefont#1{#1}\fi
\expandafter\ifx\csname citenamefont\endcsname\relax
  \def\citenamefont#1{#1}\fi
\expandafter\ifx\csname url\endcsname\relax
  \def\url#1{\texttt{#1}}\fi
\expandafter\ifx\csname urlprefix\endcsname\relax\def\urlprefix{URL }\fi
\providecommand{\bibinfo}[2]{#2}
\providecommand{\eprint}[2][]{\url{#2}}

\bibitem[{pre()}]{previous}
\bibinfo{note}{R.~Davis et al., Phys. Rev. Lett. {\bf 20}, 1205 (1968);
  R.~Becker-Szendy et al., Phys. Rev. D {\bf 46}, 3720 (1992); K.~S.~Hirata et
  al., Phys. Lett. B {\bf 280}, 146 (1992); J.~N.~Abdurashitov et al., Phys.
  Lett. B {\bf 328}, 234 (1994); M.~Sanchez et al., Phys. Rev. D {\bf 68},
  113004 (2003); M.~H.~Ahn et al, Phys. Rev. D {\bf 74}, 072003 (2006)}.

\bibitem[{sup({\natexlab{a}})}]{superksays}
\bibinfo{note}{Y.~Fukuda et al., Phys. Rev. Lett. {\bf 81}, 1562 (1998);
  Y.~Ashie et al., Phys. Rev. Lett. {\bf 93}, 101801 (2004); Y.~Ashie et al.,
  Phys. Rev. D {\bf 71}, 112005 (2005)}.

\bibitem[{sol()}]{solarprev}
\bibinfo{note}{P. ~Anselmann et al., Phys. Lett. B {\bf 285}, 376 (1992);
  J.~N.~Abdurashitov et al., Phys. Rev. Lett. B {\bf 328}, 234 (1994);
  Y.~Fukuda et al., Phys. Rev. Lett. {\bf 86}, 5651 (2001); J.~Hosaka et al.,
  Phys. Rev. D {\bf 73}, 112001 (2006); Q.~R.~Ahmad et al., Phys. Rev. Lett.
  {\bf 89}, 011301 (2006)}.

\bibitem[{kam()}]{kamland}
\bibinfo{note}{S.~Abe et al., Phys. Rev. Lett. {\bf 100}, 221803 (2008);
  T.~Araki et al., Phys. Rev. Lett. {\bf 94}, 081801 (2005)}.

\bibitem[{\citenamefont{Adamson et~al.}(2008)}]{Adamson:2007gu}
\bibinfo{author}{\bibfnamefont{P.}~\bibnamefont{Adamson}} \bibnamefont{et~al.},
  \bibinfo{journal}{Phys. Rev. D} \textbf{\bibinfo{volume}{77}},
  \bibinfo{pages}{072002} (\bibinfo{year}{2008}).

\bibitem[{lep()}]{lep}
\bibinfo{note}{M.~Acciarri {et al}., Phys. Lett. B {\bf 431}, 199 (1998);
  P.~Abreu {et al}., Z. Phys. {\bf C74}, 577 (1997); R.~Akers {et al}., Z. Phys
  {\bf C65}, 47 (1995); D.~Buskulic {et al}., Phys. Lett. B {\bf 313}, 520
  (1993); G.~S.~Abrams et al., Phys. Rev. Lett {\bf 63}, 2173 (1989)}.

\bibitem[{\citenamefont{de~Gouvea et~al.}(2007)\citenamefont{de~Gouvea,
  Jenkins, and Vasudevan}}]{deGouvea:2006gz}
\bibinfo{author}{\bibfnamefont{A.}~\bibnamefont{de~Gouvea}},
  \bibinfo{author}{\bibfnamefont{J.}~\bibnamefont{Jenkins}}, \bibnamefont{and}
  \bibinfo{author}{\bibfnamefont{N.}~\bibnamefont{Vasudevan}},
  \bibinfo{journal}{Phys. Rev. D} \textbf{\bibinfo{volume}{75}},
  \bibinfo{pages}{013003} (\bibinfo{year}{2007}), \eprint{hep-ph/0608147}.

\bibitem[{sup({\natexlab{b}})}]{supernovae}
\bibinfo{note}{G.~C.~McLaughlin et al., Phys. Rev. C {\bf 59}, 2873 (1999);
  D.~O.~Caldwell et al., Phys. Rev. D. {\bf 61}, 123005 (2000); J.~Fetter et
  al., Astropart. Phys. {\bf 18}, 433 (2003)}.

\bibitem[{\citenamefont{Ahmad et~al.}(2002)}]{Ahmad:2002jz}
\bibinfo{author}{\bibfnamefont{Q.~R.} \bibnamefont{Ahmad}}
  \bibnamefont{et~al.}, \bibinfo{journal}{Phys. Rev. Lett.}
  \textbf{\bibinfo{volume}{89}}, \bibinfo{pages}{011301}
  (\bibinfo{year}{2002}), \eprint{nucl-ex/0204008}.

\bibitem[{sup({\natexlab{c}})}]{superktau}
\bibinfo{note}{Y.~Fukuda et al., Phys. Rev. Lett. {\bf 85}, 3999 (2000); K.~Abe
  et al., Phys. Rev. Lett. {\bf 97}, 171801 (2006)}.

\bibitem[{ste()}]{steriletheory}
\bibinfo{note}{G.~L.~Fogli et al., Phys. Rev. D {\bf 64}, 093005 (2001);
  A.~Donini et al. JHEP {\bf 12}, 013 (2007); A.~Dighe and S.~Ray, Phys. Rev. D
  {\bf 76}, 113001 (2007)}.

\bibitem[{min()}]{minosCC}
\bibinfo{note}{D.~G.~Michael et al., Phys. Rev. Lett. {\bf 97}, 191801 (2006);
  P.~Adamson et al., Phys. Rev. Lett. {\bf 101} 131802 (2008)}.

\bibitem[{\citenamefont{Michael et~al.}(2008)}]{nim}
\bibinfo{author}{\bibfnamefont{D.~G.} \bibnamefont{Michael}}
  \bibnamefont{et~al.} (\bibinfo{year}{2008}), \eprint{{\it accepted to} Nucl.
  Instrum. Meth., arXiv:0805.3170 [physics.ins-det]}.

\bibitem[{\citenamefont{Tagg et~al.}(2005)}]{Tagg:2004bu}
\bibinfo{author}{\bibfnamefont{N.}~\bibnamefont{Tagg}} \bibnamefont{et~al.},
  \bibinfo{journal}{Nucl. Instrum. Meth.} \textbf{\bibinfo{volume}{A539}},
  \bibinfo{pages}{668} (\bibinfo{year}{2005}).

\bibitem[{\citenamefont{Lang et~al.}(2005)}]{Lang:2005xu}
\bibinfo{author}{\bibfnamefont{K.}~\bibnamefont{Lang}} \bibnamefont{et~al.},
  \bibinfo{journal}{Nucl. Instrum. Meth.} \textbf{\bibinfo{volume}{A545}},
  \bibinfo{pages}{852} (\bibinfo{year}{2005}).

\bibitem[{\citenamefont{Raufer}(2007)}]{tobi}
\bibinfo{author}{\bibfnamefont{T.~M.} \bibnamefont{Raufer}},
  \bibinfo{journal}{Ph.D. Thesis, Oxford University}  (\bibinfo{year}{2007}).

\bibitem[{\citenamefont{Litchfield}(2008)}]{phill}
\bibinfo{author}{\bibfnamefont{R.~P.} \bibnamefont{Litchfield}},
  \bibinfo{journal}{Ph.D. Thesis, Oxford University}  (\bibinfo{year}{2008}).

\bibitem[{\citenamefont{Osiecki}(2007)}]{osiecki}
\bibinfo{author}{\bibfnamefont{T.}~\bibnamefont{Osiecki}},
  \bibinfo{journal}{Ph.D. Thesis, University of Texas at Austin}
  (\bibinfo{year}{2007}).

\bibitem[{\citenamefont{Apollonio et~al.}(2003)}]{Apollonio:2002gd}
\bibinfo{author}{\bibfnamefont{M.}~\bibnamefont{Apollonio}}
  \bibnamefont{et~al.}, \bibinfo{journal}{Eur. Phys. J.}
  \textbf{\bibinfo{volume}{C27}}, \bibinfo{pages}{331} (\bibinfo{year}{2003}),
  \eprint{hep-ex/0301017}.

\end{thebibliography}
